\documentclass[pra,twocolumn,showpacs,floatfix,amssymb,superscriptaddress,
amsmath,fontsize=11pt]{revtex4-1}

\usepackage{times}
\usepackage{amssymb}
\usepackage{graphics} 
\usepackage{dcolumn}
\usepackage[latin1]{inputenc}
\usepackage[mathscr]{eucal}
\usepackage{epsfig} 
\usepackage{rotating} 
\usepackage{bm}
\usepackage{bbm}
\usepackage{textcomp}
\usepackage[american]{babel}

\newcommand{\be}{\begin{equation}}
\newcommand{\ee}{\end{equation}}
\newcommand{\bee}{\begin{equation*}}
\newcommand{\eee}{\end{equation*}}
\newcommand{\tvac}[1]{T^0_{#1}}
\newcommand{\vp}[1]{V_{#1}}
\newcommand{\tmb}[1]{T^{\mathrm{MB}}_{#1}}
\newcommand{\bq}{\bm{q}}
\newcommand{\bn}{\vec{n}}
\newcommand{\bk}{\bm{k}}
\newcommand{\bp}{\bm{p}}
\newcommand{\br}{\bm{r}}
\newcommand{\eps}[1]{\varepsilon_{#1}^{}}
\newcommand{\ene}[1]{\xi_{#1}^{}}
\newcommand{\omz}{\omega_z^{}}
\newcommand{\om}{\omega}
\newcommand{\lz}{\ell_z^{}}

\newcommand{\nr}{n_{r}^{}}
\newcommand{\nrp}{n_{r}'}

\newcommand{\ra}{\rangle}
\newcommand{\la}{\langle}
\newcommand{\ef}{\varepsilon_{F}^{}}
\newcommand{\efs}{\varepsilon_{F,\sigma}^{}}

\newcommand{\nf}{n_{F}^{}}

\newcommand{\as}{a_{3D}}

\newcommand{\da}{\dagger}

\begin{document}

\title{Pairing instabilities in quasi-two-dimensional Fermi gases}

\author{Ville Pietil\"a}
\affiliation{Physics Department, Harvard University, Cambridge, 
MA 02138, USA}

\author{David Pekker}
\affiliation{Department of Physics, California Institute of Technology, MC 114-36, Pasadena, 
CA 91125, USA}

\author{Yusuke Nishida}
\affiliation{Center for Theoretical Physics, Massachusetts Institute of Technology, Cambridge, MA 02139, USA}

\author{Eugene Demler}
\affiliation{Physics Department, Harvard University, Cambridge, 
MA 02138, USA}

\begin{abstract}
We study non-equilibrium dynamics of ultracold two-component Fermi gases in 
low-dimensional geometries after the interactions are quenched from a weakly 
interacting to a strongly interacting regime. We develop a $T$-matrix formalism that takes 
into account the interplay between Pauli blocking and tight confinement in low-dimensional 
geometries. We employ our formalism to study the formation of  molecules 
in quasi-two-dimensional Fermi gases near Feshbach resonance and show that the rate at 
which molecules form depends strongly on the transverse confinement. Furthermore, 
Pauli blocking gives rise to a sizable correction to the binding energy of molecules. 
\end{abstract}


\maketitle

\section{Introduction} Theoretical prediction and experimental observation of 
magnetic-field-induced 
Feshbach resonances in ultracold atoms~\cite{Bloch:2008} paved the way for many 
exciting discoveries, including demonstration of fermionic superfluidity~\cite{Zwierlein:2005},
observation of Efimov trimers and Fermi polarons~\cite{Lompe:2010,Mathy:2011,
Chevy:2006,Prokofev:2008,Schirotzek:2009}, and creation of quantum degenerate gases of 
polar molecules~\cite{Ni:2008}. Surprisingly, additional Feshbach resonances can be found in 
systems with reduced dimensionality. Earlier theoretical work on two particle scattering
in systems confined to one-dimensional tubes~\cite{Olshanii:1998} and two-dimensional 
pancakes~\cite{Petrov:2001} suggested a possibility realizing ``confinement-induced 
resonances" (CIR's), i.e., special scattering resonances made possible by restricting the transverse 
motion of atoms. Such resonances have been observed in both 
one-dimensional~(1D)~\cite{Moritz:2005} and two-dimensional~(2D)~\cite{Frohlich:2011} 
systems. In optical lattices, Feshbach resonances can 
give rise to nontrivial manifestations of mixing of higher Bloch bands \cite{Mathy:2009,Cui:2010}. 

Most of the earlier work has focused on the interplay of dimensional confinement
and resonant interactions in two-body problems. Very few extensions to many-body systems 
have been considered so far. On the other hand, the primary motivation for studying 
low-dimensional systems is to understand the surprising properties of low dimensional many-body 
systems (see Refs.~\cite{Giamarchi:2004,Petrov:2004,Cazalilla:2011} for a review). 
Moreover, experiments are always performed in systems with a finite density; in many cases 
it may not be easy to disentangle many-body effects from two-particle scattering.
For example, confinement-induced molecules are relatively large on the BCS side of 
resonance~\cite{Petrov:2001}. Already for a modest density of fermions, distances between 
particles may become comparable to the size of bound pairs, and the Pauli principle can have a 
strong effect on the collisional properties of atoms and, as a result, on the properties of CIR's.

In this paper we provide a theoretical analysis of a many-body system composed of 
two-component fermions confined in two-dimensional geometries in the vicinity of a Feshbach 
resonance. We focus on quench-type experiments, where a noninteracting mixture is rapidly 
taken to the regime of strong interactions~\cite{Jo:2009,Pekker:2011}. We analyze many-body 
corrections to the energies of confinement-induced molecules and calculate the rate at which they 
are formed out of unbound atoms. 

One of the intriguing questions raised by recent experiments 
concerns the possibility of using fermionic systems close to Feshbach resonances for 
exploring many-body phenomena associated with strong repulsive interactions.
For example, positive scattering length on the BEC side of the Feshbach resonance has been
suggested as a route to observe the Stoner instability~\cite{Duine:2005}.  While the
first experiments by Jo {\it et al}.~have been interpreted using a simple
mean-field picture of such transition~\cite{Duine:2005,LeBlanc:2009}, 
subsequent measurements showed that the system is strongly dominated by 
fast molecule formation~\cite{Sanner:2011}, as predicted theoretically 
in Refs.~\cite{Pekker:2011,Zhang:2011}. In this paper, we demonstrate that dimensional confinement can have a dramatic effect on the dynamics of molecule formation. We find that the peak in the molecule formation rate should be shifted from the BEC to the
BCS side of the resonance with increasing transverse confinement. Testing our predictions in 
experiments will help to distinguish between different models of molecule 
formation~\cite{Pekker:2011,Zhang:2011}.

Another conceptually intriguing aspect of the system we study is that one cannot use separation 
of energy scales to simplify the analysis. Typically when many-body systems of ultracold atoms 
are studied, it is assumed that one can start by solving a two-body problem to obtain the strength 
of contact interaction and then work with this contact interaction when analyzing the many-body 
problem. In our system the effective two particle scattering can be strongly modified by the 
presence of other particles~\cite{Pekker:2011}. Hence, an accurate analysis of our system 
requires understanding of the interplay between few-body and many-body phenomena.

\section{Vacuum $T$-matrix} 
Traditionally, two-body problems in low-dimensional 
geometries have been analyzed using the Schr\"odinger equation, which can be simplified 
into two decoupled single-particle problems corresponding to the relative and the center-of-mass 
(c.m.) motion~\cite{Olshanii:1998,Petrov:2001}. This approach is, however, difficult to 
generalize to the many-body case. In the presence of a filled Fermi sea, the c.m. momentum 
of the scattering pair relative to the Fermi sea is important and cannot be taken into account by a 
simple momentum boost. Therefore we re-examine the two-body problem in 
quasi-2D geometries by recasting the results of Ref.~\cite{Petrov:2001} to the form of a $T$-matrix 
in vacuum. For a discussion regarding Feshbach resonances in low-dimensional systems, 
see Ref.~\cite{Kestner:2007}. We take the gas to be homogeneous in a 2D plane and assume a 
strong harmonic confinement in the transverse direction. 

We start from the full 3D scattering problem and use a contact interaction 
$V_{\mathrm{int}}^{}(\br-\br') = V_0\delta(\br-\br')$ to describe the inter-particle interactions. 
In order to make the connection to the many-body problem we do not separate relative and 
center-of-mass motion from the outset. This gives rise to a $T$-matrix which depends on  
energy $\hbar\omega$ as well as on the harmonic oscillator 
quantum numbers $\bn = (n_1^{},n_2^{})$  and $\bn' = (n_1',n_2')$ corresponding to incoming 
and outgoing particles. For the contact interaction the Lippmann-Schwinger equation 
takes a simple form
\be
\label{tvac}
\tvac{\bn,\bn'}(\om) = \vp{\bn,\bn'} + \sum_{\bn''}\vp{\bn,\bn''}\Pi^{(0)}_{\bn''}(\om)\,
\tvac{\bn'',\bn'}(\om),
\ee
where the polarization operator is given by 
 \be
\label{vac_pol}
\Pi^{(0)}_{\bn}(\om) = \int\frac{d\bk}{(2\pi)^2}\,\frac{1}{\hbar\om-2\eps{\bk}-
\hbar\om_z^{}(n_1^{}+n_2^{}) + i0^+_{}},
\ee
and we have denoted the trap frequency in transverse direction by $\om_z^{}$. The polarization 
operator in vacuum has the property $\Pi^{(0)}_{\bn}(\om) = \Pi^{(0)}_{n_1^{} + n_2^{}}(\om)$ for 
$\bn = (n_1^{},n_2^{})$. We will utilize these two notations interchangeably when discussing the 
properties of the many-body $T$-matrix. The dispersion 
is given by  $\eps{\bk} = \hbar^2 k^2/2m$, and we measure energies and frequencies with 
respect to the zero-point energy $\hbar\omega_z$. Thus, a confined particle 
in the lowest vibrational state with no in-plane momentum is assumed to have zero energy.

Since $V_{\mathrm{int}}^{}$ depends only on the relative motion of scattering particles, 
we write the matrix elements $\vp{\bn,\bn'}$ in terms of the quantum numbers corresponding 
to relative ($\nr$) and center-of-mass motion ($N$)
\be
\label{matrix_elements}
\vp{\bn,\bn'} = V_0\sum_{N,\nr,\nrp}\mathcal{C}_{N,\nr}^{\bn^{\scriptstyle\, *}}
\mathcal{C}_{N,\nrp}^{\bn'} \varphi_{\nr}^*(0)\,\varphi_{\nrp}^{}(0).
\ee
Here
$\varphi_{\nr}^{}$ is a harmonic oscillator eigenfunction corresponding 
to  relative motion and the harmonic oscillator length in the transverse direction is denoted by 
$\lz=\sqrt{\hbar/m\omz}$. The Clebsch-Gordan coefficients arising from the change of basis 
are defined as  $\mathcal{C}_{N,\nr}^{\bn} = \la N,\nr|n_1^{},n_2^{} \ra$. Quantum numbers 
$n_1^{},n_2^{},\nr $, and $N$ are non-negative integers and energy conservation imposes 
condition $n_1^{} + n_2^{} = N + \nr$ for the non-zero elements $\mathcal{C}_{N,\nr}^{\bn}$.

The form of matrix elements $\vp{\bn,\bn'}$ suggests we look for a solution in the basis of 
relative and c.m. quantum numbers and then go back to the original basis. We find that 
(see Appendix~A)
\begin{align}
\label{vacuum_solution}
\tvac{\bn,\bn'}(\om) = \sqrt{2\pi}\,\ell_z^{}\sum_{N,\nr,\nrp}\mathcal{C}_{N,\nr}^{\bn^{\scriptstyle*}} 
&\mathcal{C}_{N,\nrp}^{\bn'}\,\varphi_{\nr}^*(0)\varphi_{\nrp}^{}(0) \notag \\
&\times\,\mathcal{T}_{0}(\om-N\omega_z^{}).
\end{align}
The structure of $\tvac{\bn,\bn'}$ shows explicitly the 
decoupling of relative and c.m. motion. Furthermore, since the interaction potential depends 
only on the relative motion, the c.m. quantum number does not change in the scattering and 
contributes only as shift to the energy of scattering particles. When the bare interaction 
$V_0^{}$ is eliminated, $\mathcal{T}_0$ is given by~\cite{Petrov:2001,Lim:2008} 
(for details, see Appendix~A)
\be
\label{tvac_2D}
\frac{1}{\mathcal{T}_{0}(\om)} = \frac{m}{4\pi\hbar^2}\big[\sqrt{2\pi}\,\lz/a_{3D}^{} + 
w(\om/\omz + i0^+_{})\big],
\ee
where function $w(z)$ is defined as 
\be
\label{w_func}
w(z) = \lim_{n\rightarrow\infty}\left[2\sqrt{\frac{n}{\pi}}\ln\frac{n}{e^2}-
\sum_{\ell=0}^{n}\frac{(2\ell-1)!!}{(2\ell)!!}\ln(\ell - z/2)\right].
\ee
The double factorial is given by $n!!\equiv n\cdot(n-2)\cdot(n-4)...$, and by definition 
$(-1)!! = 0!! = 1$. The two-body $T$-matrix has a series of poles corresponding to different values 
of the center-of-mass quantum number $N$. In particular, there is a bound state  
corresponding to $N=0$ which exists for all $a_{3D}^{}$ and coincides with the Feshbach 
molecule deeply on the BEC side. Deeply on the BCS side of resonance ($|a_{3D}^{}| \ll \lz $), 
the energy of the confinement-induced two-body bound state has a simple 
expression $\varepsilon_b^{} = -\frac{B}{\pi}\hbar\omz\,e^{-\sqrt{2\pi}\,\lz/|a_{3D}^{}|}$, where 
$B=0.905$~\cite{Bloch:2008}. In general the pole has to be computed numerically 
from Eq.~\eqref{tvac_2D}.

\section{Many-body $T$-matrix and Cooperon}

Let us next discuss the many-body effects in the formation of confinement-induced molecules 
in quasi-2D geometries. The system is described by a many-body 
Hamiltonian
\begin{align}
\label{hami}
 H = &\sum_{\bk,n,\sigma} \ene{\bk,n,\sigma}\,c_{\bk,n,\sigma}^{\da}c_{\bk,n,\sigma}^{}  \notag \\ 
 &+ \frac{1}{\mathcal{V}}\sum_{\bk,\bq,\bp}\sum_{\bn,\bn'}\vp{\bn,\bn'} \,c_{\bk+\bq,n_1^{},\uparrow}^{\da}c_{\bp-\bq,n_2^{},\downarrow}^{\da}c_{\bp,n_2',\downarrow}^{}c_{\bk,n_1',\uparrow}^{},
 \end{align}
where $\ene{\bk,n_i^{},\sigma} = \eps{\bk} - \efs + \hbar\omz n_i^{}$ and particles carry $2D$ 
momentum $\bk$ as well as harmonic oscillator quantum number $n_i^{}$. We have also 
allowed a possible imbalance between the two fermion species. 

To incorporate the Pauli blocking to our analysis, we derive a $T$-matrix in the presence of 
Fermi sea (Cooperon). We approximate the full Bethe-Salpeter equation by taking into account 
the ladder diagrams and obtain
\be
\label{mb_ls}
\tmb{\bn,\bn'}(\om,\bq) = \vp{\bn,\bn'} + \sum_{\bn''}\vp{\bn,\bn''}\Pi^{}_{\bn''}(\om,\bq)\,
\tmb{\bn'',\bn'}(\om,\bq),
\ee
where we assume that the scattering particles can have finite c.m. momentum $\bq$ in the 2D 
plane. The full polarization operator $\Pi^{}_{\bn}(\om,\bq)$ is of the form
\be
\label{full_pol}
\Pi^{}_{\bn}(\om,\bq) = \int\frac{d\bk}{(2\pi)^2}\,\frac{1-\nf(\ene{\bk+\bq,n_1^{},\uparrow}) -
\nf(\ene{\bk,n_2^{},\downarrow})}{\hbar\om(1+i0^+)-\ene{\bk+\bq,n_1^{},\uparrow}-\ene{\bk,n_2^{},\downarrow}}.
\ee

Although c.m. and relative motion become coupled in the presence of Fermi sea, we utilize 
insights from the two-body problem and look for a solution where c.m. and relative motion 
are at least partially decoupled. We find that the solution of Bethe-Salpeter equation can be 
written in terms of a $T$-matrix depending only on the c.m. quantum numbers
\be
\label{mb_solution}
\tmb{\bn,\bn'} = \sqrt{2\pi}\,\ell_z^{}\sum_{\substack{N,N' \\ \nr,\nrp}}\mathcal{C}_{N,\nr}^{\bn^{\scriptstyle\,*}}\,
\mathcal{C}_{N',\nrp}^{\bn'}\,
\varphi_{\nr}^*(0)\varphi_{\nrp}^{}(0)\,\mathcal{T}_{N,N'}^{}.
\ee
We use the two-body $T$-matrix $\mathcal{T}_0^{}$ to renormalize 
the UV divergence associated with polarization operator~\eqref{full_pol} and obtain 
(for details, see Appendix~B)
\be
\label{t_com}
\mathcal{T}^{-1}_{N,N'}(\om,\bq) = \mathcal{T}^{-1}_{0}(\om-N\omega_z^{}-\omega_{\bq}^{})
\delta_{N,N'}^{}  - \mathcal{D}_{N,N'}^{}(\om,\bq),
\ee
where the renormalized polarization operator is given by 
\begin{align}
\label{dop}
\mathcal{D}_{N,N'}^{} =  \sum_{\bn,\nr} & \,u_{\nr,N+\nr-K}^{}  \,
\mathcal{C}_{K,N+\nr-K}^{\bn^{\scriptstyle\,*}}\,\mathcal{C}_{N,\nr}^{\bn} \notag \\
& \quad\quad \times  [\Pi_{\bn}^{}(\om,\bq)-\Pi_{\bn}^{(0)}(\om-\omega_{\bq}^{})].
\end{align}
The coefficients $u_{n,m}^{}$ are related to the zeros of the harmonic oscillator eigenfunctions 
[see Eq.~\eqref{hoEigFunc}] and they are given by
\be 
\label{coeffs}
u_{n,m}^{} = 
\frac{(-1)^{(n+m)/2}(n-1)!!\,(m-1)!!}{\sqrt{n!\,m!}}
\ee 
for even and non-negative $n$ and $m$. Otherwise $u_{n,m}^{} $ is zero.
We have also defined $\hbar\omega_{\bq}^{} = 
\frac{1}{2}\varepsilon_{\bq}^{}-\varepsilon_{F,\uparrow}^{}-\varepsilon_{F,\downarrow}^{}$. 
In order to correctly renormalize the UV divergence associated with the 
2D momentum integral in Eq.~\eqref{full_pol}, we have to evaluate the two-body $T$-matrix such 
that  the Fermi surface and finite c.m. momentum are taken into account. This shifts the argument 
of $\mathcal{T}^{}_{0}$ by $\omega_{\bq}^{}$ in Eq.~\eqref{t_com}.

Conservation of energy and parity impose selection rules for the allowed scattering processes 
and render matrix  $V_{\bn,\bn'}^{}$ non-invertible. Since both $T^0$ and $T_{}^{MB}$ share 
the same structure as $V_{\bn,\bn'}^{}$, they also lack well-defined inverses and 
Eqs.~\eqref{vacuum_solution} and~\eqref{mb_solution} have to be solved in terms of matrices 
$\mathcal{T}_0^{}$ and $\mathcal{T}$ which are both regular. The full solution retains 
all discrete energy levels in the transverse direction and although the most 
interesting 2D limit does not involve real processes via higher bands, virtual scattering processes 
become important near the Fesh\-bach resonance. The general solution based on 
Eqs.~\eqref{mb_solution}--\eqref{dop} enables a systematic 
analysis of pairing instabilities from the strictly 2D regime at zero temperature to the 
confinement dominated 3D regime where temperature and Fermi energy become 
comparable with $\hbar\omega_z^{}$.  

\section{Molecule formation} 

To analyze the possible pairing instabilities, we assume that the 
system is initially spin balanced and weakly interacting. In the spirit of  Ref.~\cite{Jo:2009}, we 
consider an instantaneous quench where interactions are rapidly 
modified utilizing a 3D Feshbach resonance. 
The molecule formation is associated with the appearance of poles 
$\hbar\om = \Omega_{\bq} + i\Delta_{\bq}^{}$ in the many-body $T$-matrix 
$\tmb{\bn,\bn'}(\om,\bq)$~\cite{Randeria:1989}. We identify the real part $ \Omega_{\bq}$ as 
the binding energy of the molecule and the imaginary part $\Delta_{\bq}^{}$ as the growth rate of the instability toward formation of molecules~\cite{Pekker:2011}. 

\begin{figure}[h!]
\centering
\includegraphics[width=0.475\textwidth]{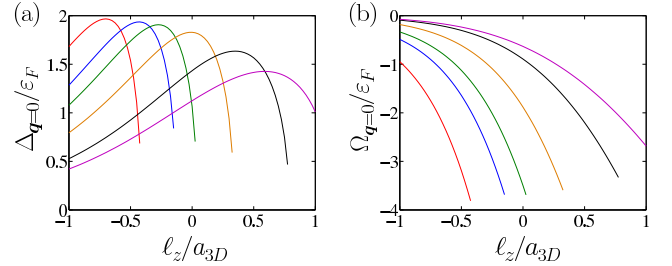}
\caption{\label{peaks}  (Color online) Growth rate of the pairing instability (a) and the binding 
energy of molecules (b)  at zero temperature as a function of $\lz/\as$. The values of  
$\ef/\hbar\omz$ are (from left to right) $\ef/\hbar\omz = 0.0175,\,0.025,\,0.0375,\,0.075,\,0.2$, and 
$\,0.4$. } 
\end{figure}

Similarly to the 3D case~\cite{Pekker:2011}, we find that the system exhibits an instability 
towards molecule formation via two-body processes as long as the Fermi sea can absorb the 
binding energy of the molecules. This results in a sharp cutoff in the growth rate, see  
Fig.~\ref{peaks}(a). For a fixed $\lz/a_{3D}^{}$, the binding energy of molecules depends 
strongly on the ratio $\ef/\hbar\omz$ and Fig.~\ref{peaks}(b) shows that the binding energy 
increases with increasing strength of the transverse confinement. The location of the peak value 
for the growth rate of instability can be varied by adjusting the ratio $\lz/a_{3D}^{}$ and, 
in particular, tight enough transverse confinement can move the pairing instability completely to  
the BCS side. On the other hand, when $\ef/\hbar\omega_z^{}\simeq 0.1$ as in 
Refs.~\cite{Jochim:2003a,Frohlich:2011}, the pairing instability extends to the BEC side and 
fast two-body processes dominate the three-body processes. When the molecule formation via 
two-body processes is no longer possible, the leading instability is a three-body recombination 
which is suppressed for Fermi gases due to low densities and the Pauli principle.

\begin{figure}[h!]
\centering
\includegraphics[width=0.325\textwidth]{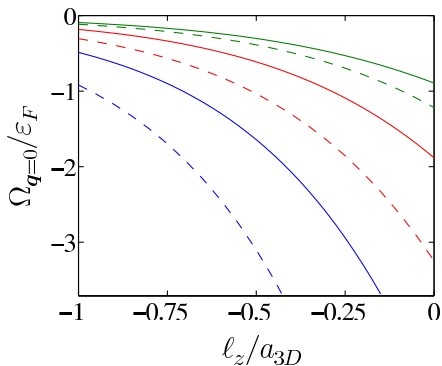}
\caption{\label{binding_energy}  (Color online) Binding energy of molecules at zero 
temperature for $\ef/\hbar\omz=0.025,\,0.075$, and $0.2$ (from 
bottom to top).  The binding energy in the presence of Fermi sea (solid lines) is always smaller 
than the vacuum binding energy (dashed lines) for the parameters investigated. 
} 
\end{figure}

The binding energy in vacuum is compared to the binding energy at finite densities in 
Fig.~\ref{binding_energy}. The relation between vacuum and finite density binding energies 
depends again on $\ef/\hbar\omz$, and when the system becomes more 
three-dimensional~(3D)  
(i.e.~when $\ef/\hbar\omz$ increases), Pauli blocking by the Fermi sea can result 
in a stronger binding of molecules. The crossover takes place roughly at $\ef/\hbar\omz = 0.5$. 
In 3D gases many-body corrections always result in stronger binding~\cite{Pekker:2011} 
and Pauli blocking induced weaker binding is a manifestation of  2D physics.  
On the other hand, the binding energy of the molecules is larger than the binding energy of the 
Feshbach molecules existing on the BEC side of resonance. 

\begin{figure}[h!]
\centering
\includegraphics[width=0.325\textwidth]{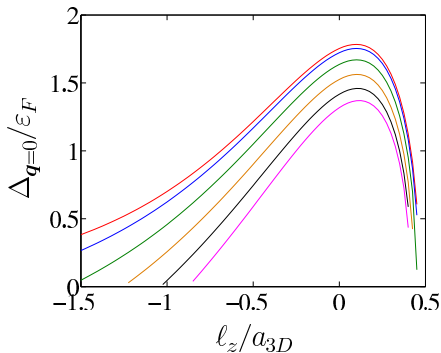}
\caption{\label{damping}  (Color online) The growth rate of the pairing instability as a function 
of $\lz/\as$ for $\ef/\hbar\omz =0.1$. The temperature is (from top to bottom) 
$T/T_F^{}=0,\, 0.1,\,0.2,\,0.3,\,0.4$, and $0.5$. } 
\end{figure}

Finite temperature suppresses strongly the growth rate of pairing instability whereas the 
binding energy decreases more slowly with increasing temperature. In Fig.~\ref{damping}, 
the growth rate is shown at different temperatures for $\ef/\hbar\omz=0.1$ corresponding  
to the experimental parameters of Refs.~\cite{Jochim:2003a,Frohlich:2011}. 
Pairing instability at the BCS side of the resonance is sensitive to the temperature since 
thermal fluctuations can easily break molecules at small binding energies. At high enough 
temperatures the pairing instability can become completely suppressed for weak attractive 
interactions. On the other hand, the cutoff in the growth 
rate at $\ell_z^{}/\as \approx 0.5$ does not in general depend strongly on the temperature. 
We note that although the pairing instability can persist to quite high temperatures, the critical 
temperature for the superfluid transition is typically much lower near the Feshbach resonance 
or deeply in the BEC regime~\cite{Petrov:2003,Iskin:2009}. 

The vacuum $T$-matrix has several poles on the BEC side of the Feshbach resonance 
corresponding to the different c.m. quantum numbers. This can in principle give rise to 
se\-ve\-ral pairing instabilities which show up as distinct poles in the Cooperon. For small 
$\ef/\hbar\omz $ these 
poles do not coexist for given $\lz/a_{3D}^{}$ since the Fermi sea is unable to absorb large 
binding energies. When $\ef/\hbar\omz $ increases, the poles start to overlap and additional 
poles with non-zero $\Delta_{\bq}$ become discernible. However, these additional instabilities 
remain weak compared to the primary instability corresponding to the pole of the vacuum 
$T$-matrix with $N=0$.  

So far we have analyzed the pairing instability in the case of zero c.m. momentum $\bq$. 
We find that the results remain qualitatively the same for finite $\bq$ and in the spin-balanced 
case the most unstable mode is always at $\bq=0$. However, the growth rate of instability 
decreases slowly as a function of $|\bq|$; in a realistic quench 
experiment it is likely that molecules with a wide distribution of momenta are created. We find 
that finite $\bq$ reduces the binding energy due to the smaller 
number of low-energy states that are available for scattering~\cite{Schreiffer:1964}. In 
spin-imbalanced systems, the lowest energy state can shift to finite 
momentum~\cite{Zoellner:2011,Parish:2011}.

\section{Discussion} 
We have studied pairing instabilities in spin balanced quasi-2D 
Fermi gases when interactions are dynamically quenched to the regime of strong interactions 
using 3D Feshbach resonances.
We found that the pairing instability can be shifted to the BCS side of resonance by adjusting 
the axial confinement with respect to Fermi energy. Pauli blocking was found to 
renormalize significantly the binding energies, resulting in weaker binding in the 2D limit than 
that warranted by the two-body description. 

The growth rate of pairing instability can be measured by monitoring the atom 
loss~\cite{Jo:2009} and the binding energy can be probed using rf 
spectroscopy~\cite{Schunck:2007,Frohlich:2011,Sommer:2011}.
In a related work~\cite{rf_paper}, we argue that the recent experiment~\cite{Frohlich:2011} 
probing the properties of 2D Fermi gases can be interpreted in terms of dynamically created 
polarons.  Another recent experiment~\cite{Sommer:2011} measures directly the binding energies 
of the molecules and finds agreement with a theoretical prediction for the two-body bound states 
in 1D optical lattices~\cite{Orso:2005}. On the other hand, our calculation 
(Fig.~\ref{binding_energy}) predicts that the two-body bound-state energy should be significantly 
renormalized by the presence of the Fermi sea. The discrepancy could stem from the fact that 
our calculation probes an unpaired gas which is rapidly quenched to the strongly interacting 
regime, whereas in Ref.~\cite{Sommer:2011} the system corresponds to a strongly interacting 
gas in equilibrium with a large number of paired atoms.

The $T$-matrix approach presented here can be used to probe the competition between 
polaron and molecule~\cite{Zoellner:2011,Klawunn:2011,Parish:2011,Combescot:2007} in 
quasi-2D systems and to investigate dimensional crossover from 
2D to 3D~\cite{Iskin:2009,Bertaina:2011,Dyke:2011,Sommer:2011}.  
Our formalism is also useful for studies of pair formation in other low-dimensional geometries 
and Bose gases. In particular, it can be used to investigate the effective three-body collisions induced by virtual excitations of the transverse 
modes~\cite{Mazets:2008,Tan:2010,Mazets:2010}.

\begin{acknowledgements}
We would like to thank L.~Radzihovsky, M.~K\"ohl, and W.~Zwerger 
for insightful  discussions. This work was supported by the Academy of Finland (VP), the 
Army Research Office with funding from the DARPA OLE program, Harvard-MIT CUA, 
NSF Grant No. DMR-07-05472, AFOSR Quantum Simulation MURI, AFOSR MURI on Ultracold  
Molecules, and the ARO-MURI on Atomtronics.
\end{acknowledgements}

\appendix

\section{Vacuum $T$-matrix}
We briefly discuss the technical details regarding the calculation of the vacuum $T$-matrix as well 
as the many-body $T$-matrix. For simplicity, we set $\hbar=1$ in Appendices~A and~B.

The form of the matrix elements in Eq.~\eqref{matrix_elements} suggests the following ansatz 
for the vacuum $T$-matrix
\begin{equation}
\label{ansatz1}
T_{\vec{n},\vec{n}'}^0(\om)  = \sqrt{2\pi}\ell_z^{}\sum_{N,n_r^{}, n_r'} C^{\vec{n}^{\scriptstyle\, *}}_{N,n_r} C^{\vec{n}'}_{N,n_r'} \varphi_{n_r}^*(0) \varphi_{n_r'}(0)\mathcal{T}_N^{}(\om).
\end{equation}
Substituting Eq.~\eqref{ansatz1} into Eq.~\eqref{tvac} we obtain
\begin{align*}
\mathcal{T}_N^{}(\om) &= \frac{V_0}{\sqrt{2\pi}\ell_z^{}} + \notag \\
&\sum_{n_r} \int \frac{d\bk}{(2\pi)^2} \frac{  \sqrt{2\pi}\ell_z^{}|
 \varphi_{n_r}(0)|^2 V_0}{\om- \frac{k^2}{m}- \omega_z^{}(N+n_r)+i0^+_{}} \,\mathcal{T}_N^{}(\om).
\end{align*}
We observe that the c.m. index comes only through the shift of energy. Thus we can take 
$T_N(\om) =T_0(\om-N\omega_z^{})$
and for $T_0(\om)$ we obtain
\begin{align}
\label{TwoParticleT0}
\frac{1}{T_0(\om)} = \frac{\sqrt{2\pi}\ell_z^{}}{V_0} - \sum_{n_r} \int \frac{d\bk}{(2\pi)^2} \frac{ \sqrt{2\pi}\ell_z^{} 
| \varphi_{n_r}(0)|^2}{\om- \frac{k^2}{m}- \omega_z^{} n_r+i0^+_{}}.
\end{align}

We can calculate the integral in Eq.~\eqref{TwoParticleT0} using the identity
\be
\label{trick}
\frac{1}{A} = - \int_0^{\mp\infty} d\tau\, e^{A\tau},
\ee
where $\mp = \mathrm{sgn}(\mathrm{Re}\,A)$. For Eq.~\eqref{TwoParticleT0} we have two cases: 
(a) $\mathrm{Re}\,\om < 0$ and (b) $\mathrm{Re}\,\om \geq 0$.  We discuss  
case (a), and case (b) follows from an analogous calculation. 
We note that the harmonic oscillator eigenfunctions in 
Eq.~\eqref{TwoParticleT0} satisfy
\be
\label{hoEigFunc}
\sqrt{2\pi}\ell_z^{} | \varphi_{n}(0)|^2 = 
\begin{cases}
\frac{(n-1)!!}{n!!}, & \text{for even}\, n \\
0, & \text{for odd}\, n.
\end{cases}
\ee
Using the identity in Eq.~\eqref{trick}, we obtain
\begin{align*}
I & =  \sum_{n_r} \int \frac{d\bk}{(2\pi)^2} \frac{ \sqrt{2\pi}\ell_z^{} 
| \varphi_{n_r}(0)|^2}{\om- \frac{k^2}{m}- \omega_z^{} n_r+i0} \notag \\
& =  -\int_0^\infty d\tau\,\sum_{n=0}^{\infty}\frac{(2n-1)!!}{(2n)!!}\,e^{\tau(\om-2n\omega_z^{}
+i0^+_{})}_{}\,\left(\frac{m}{4\pi\tau}\right) \notag \\
& = - \int_0^\infty d\tau\,e^{\tau(\om+i0^+_{})}_{}\left( \frac{e^{\omega_z^{}\tau}_{}}{2\sinh \omega_z^{}\tau}\right)^{1/2}
\left(\frac{m}{4\pi\tau}\right).
\end{align*}
The c.m. part of the quasi-2D $T$-matrix satisfies, therefore, an equation
\be
\label{2d}
\frac{1}{T_0^{}(\om)} = \frac{\sqrt{2\pi}\ell_z^{}}{V_0} + \int_0^\infty d\tau\,e^{\tau(\om+i0^+_{})}_{}
\sqrt{\frac{e^{\omega_z^{}\tau}_{}}{2\sinh \omega_z^{}\tau}}\left(\frac{m}{4\pi\tau}\right).
\ee

The UV divergence associated with the original contact interaction is manifested as a singularity 
of the integrand in the limit $\tau\rightarrow 0$. We regularize this  divergence using the 3D 
$T$-matrix, which is given by an analogous equation:
\be
\label{3d}
\frac{1}{T_{3D}^{}(\om)} = \frac{1}{V_0} + \int_0^\infty d\tau\,e^{\tau(\om+i0^+_{})}_{}\,
\left(\frac{m}{4\pi\tau}\right)^{3/2}_{}.
\ee
We take $\om\rightarrow 0$ of Eq.~\eqref{3d} to obtain
\be
\frac{m}{4\pi a_{3D}^{}} = \frac{1}{V_0} + \int_0^\infty d\tau\,
\left(\frac{m}{4\pi\tau}\right)^{3/2}_{}.
\ee
Using this identity, we eliminate the bare interaction 
$V_0$ from Eq.~\eqref{2d}. This gives us an $T_0(\om)$ which is manifestly free from UV divergences:
\begin{widetext}
\be
\label{t0_final}
\frac{1}{T_0^{}(\om)} = \frac{m}{4\pi}\left\{ \frac{\sqrt{2\pi}\ell_z^{}}{a_{3D}^{}} + \int_0^\infty dx\,\frac{1}{x}\,
\bigg[e^{x(\om/\omega_z^{} + i0^+_{})}_{}\left(\frac{e^{x}_{}}{2\sinh x}\right)^{1/2} - 
\left(\frac{1}{2x}\right)^{1/2}\bigg]
\right\}.
\ee
\end{widetext}
 The latter term in Eq.~\eqref{t0_final} is the integral representation of the function $w(\om/\omega_z^{} + i0^+_{})$ defined 
 in Eq.~\eqref{w_func}. 
 
 \section{Many-body $T$-matrix}
 To solve the Bethe-Salpeter equation~\eqref{mb_ls} we generalize the ansatz in 
 Eq.~\eqref{ansatz1} and assume that the many-body $T$-matrix is of the form
 \be
\label{ansatz2}
\tmb{\bn,\bn'} = \sqrt{2\pi}\,\ell_z^{}\sum_{\substack{N,N' \\ \nr,\nrp}}\mathcal{C}_{N,\nr}^{\bn^{\scriptstyle\,*}}\,
\mathcal{C}_{N',\nrp}^{\bn'}\,
\varphi_{\nr}^*(0)\varphi_{\nrp}^{}(0)\,\mathcal{T}_{N,N'}^{},
\ee
where we have temporarily suppressed the frequency and momentum arguments. The 
polarization operator satisfies the following useful identity:
\begin{align}
\label{id}
\sum_{\vec{n}\,\vec{n}'} &\mathcal{C}_{N,\nr}^{\vec{n}'} 
\mathcal{C}_{N',\nrp}^{\vec{n}^{\scriptstyle\,*}}
\Pi^{}_{\vec{n},\vec{n}'} (\om,\bk)  =   \delta_{N,N'} \delta_{\nr,\nrp} \,
\Pi^{(0)}_{N+n_r^{}} (\om-\om_{\bk}^{})  + \notag \\
&\sum_{\vec{n}\,\vec{n}'} C_{N,\nr}^{\vec{n}'}C_{N',\nrp}^{\vec{n}^{\scriptstyle\,*}}
\big[{\Pi}^{ }_{\vec{n},\vec{n}'} (\om,\bk) - \delta_{\vec{n},\vec{n}'}^{}\Pi^{(0)}_{\vec{n}} (\om-\om_{\bk}^{})\big],
\end{align}
where $\om_{\bk}^{} = \frac{1}{2}\eps{\bk}-\varepsilon_{F,\uparrow}^{}-
\varepsilon_{F,\downarrow}$. 
Substituting the ansatz~\eqref{ansatz2} to the Bethe-Salpeter equation~\eqref{mb_ls} and using the above identity, we obtain an equation for the c.m. part:
\begin{align}
\label{matrix_eq}
\mathcal{T}_{N,N'} = \frac{V_0}{\sqrt{2\pi}\ell_z^{}} + \frac{V_0}{\sqrt{2\pi}\ell_z^{}}\sum_{K}
(\mathcal{D}_0)_{N,K}^{}\mathcal{T}_{K,N'} + \notag \\
\frac{V_0}{\sqrt{2\pi}\ell_z^{}}\sum_K(\mathcal{D})_{N,K}^{}\mathcal{T}_{K,N'}, 
\end{align} 
where matrices $\mathcal{D}_0$ and $\mathcal{D}$ are given by 
\be
\label{da}
(\mathcal{D}_0^{})_{N,K}^{} = \delta_{N,K}\sum_{\nr}u_{\nr,\nr}^{}
\Pi^{(0)}_{K+\nr}(\om-\om_{\bk}^{}), 
\ee
and Eq.~\eqref{dop}, respectively. Coefficients $u_{\nr,\nr}^{}$ are given by Eq.~\eqref{coeffs} 
in the main text. Equation~\eqref{matrix_eq} for $\mathcal{T} = (\mathcal{T}_{N,N'}^{})$ can be 
written in a matrix form
\be
\mathcal{T}^{-1} = \frac{\sqrt{2\pi}\ell_z^{}}{V_0} - \mathcal{D}_0 - \mathcal{D}.
\ee
Denoting $\mathcal{T}^{(0)} = \mathrm{diag}(\mathcal{T}_N^{})$, where 
$\mathcal{T}_N(\om) =\mathcal{T}_0(\om-N\omega)$, we obtain 
\be
\mathcal{T}^{(0)^{-1}} = \frac{\sqrt{2\pi}\ell_z^{}}{V_0} - \mathcal{D}_0.
\ee
This gives us an equation for the many-body $T$-matrix such that the UV divergence associated with 
$\mathcal{D}_0$ is renormalized
\be
\label{final}
\mathcal{T}^{-1} = \mathcal{T}^{(0)^{-1}}  - \mathcal{D}.
\ee
Equation~\eqref{final} is illustrated in more detail in Eqs.~\eqref{t_com} and~\eqref{dop} of the 
main text.

\bibliography{manuscript} 

\end{document}